\documentclass[a4paper,11pt]{article}
\usepackage{jheppub}

\usepackage{bbold}
\usepackage{amsmath,amsfonts,amsmath,mathtools,bbm,bm}
\usepackage{graphicx}
\usepackage{braket}
\usepackage{enumitem}
\usepackage{xcolor}
\usepackage[normalem]{ulem}
\usepackage{float}

\setcitestyle{numbers,square}

\def\Re{\text{Re}}
\def\({\left (}
\def\){\right )}

\graphicspath{{Figure/PNG/}{Figure/PDF/}{Figure/EPS/}{Figure/TEX/}{Figure/}}

\newcommand\mydots{\hbox to 1em{.\hss.\hss.}}

\title{Quantum bounds on the generalized Lyapunov exponents}

\author[a]{Silvia Pappalardi,}
\author[a]{Jorge  Kurchan}
\affiliation[a]{Laboratoire de Physique de l’\'Ecole Normale Sup\'erieure, ENS, Universit\'e PSL, CNRS, Sorbonne Universit\'e, Universit\'e de Paris, F-75005 Paris, France}
\emailAdd{silvia.pappalardi@phys.ens.fr}

\abstract{
We discuss the generalized quantum Lyapunov exponents $L_q$, defined from the growth rate of the powers of the square commutator. 
They may be related to an appropriately defined thermodynamic limit of the spectrum of the commutator, which plays the role of a large deviation function, obtained from the exponents 
$L_q$ via a Legendre transform.
We show that such exponents obey a generalized bound to chaos due to the fluctuation-dissipation theorem, as already discussed in the literature. The bounds for larger $q$ are actually stronger, placing a limit on the  large deviations of chaotic properties.
Our findings at infinite temperature are exemplified by a numerical study of the kicked top, a paradigmatic model of quantum chaos.
}

\begin{document} 
\maketitle
\flushbottom

\tableofcontents

\section{Introduction}

Classical chaos is well understood from the sensitivity of the dynamics to respect small changes in the initial conditions, the so-called butterfly effect. This is quantified by the Lyapunov exponent, the rate at which nearby trajectories separate exponentially in time. 
In the past few years, there has been a lot of attention to quantum chaos and in particular, on the \emph{quantum Lyapunov exponent} $\lambda_L$, defined from the intermediate-time exponential growth of the following square-commutator
\begin{equation}
    \label{sq}
    \langle \left |[A(t), B]\right |^2 \rangle \sim \epsilon^2 e^{\lambda_L t} \ ,
\end{equation}
where $\epsilon$ is a small parameter \cite{larkin1969quasiclassical}.
The interest in this object comes from the fact that $\lambda_L$ obeys a bound 
\begin{equation}
    \label{eq:bL}
    \lambda_L \leq \frac{2 \pi}{\beta \hbar } \ .
\end{equation}
This result, now known as the \emph{quantum bound to chaos}, was proved within the high energy community \cite{maldacena2016bound}, driven to the topic because maximal chaos is attained by models of Black Holes, including the Sachdev-Ye-Kitaev model (SYK) \cite{sachdev1993gapless, KitaTalk, chowdhury2021sachdev}. 
The interest in these issues has later spread over different communities, from condensed matter to quantum information theory. 
Recently, the existence of the bound in Eq.\eqref{eq:bL} has been physically rationalized as a consequence of the fluctuation-dissipation theorem (FDT) \cite{tsuji2018bound, pappalardi2022quantum} since the out of time-order correlators (OTOC) appearing in Eq.\eqref{sq} can be mapped into two-time correlation functions in a duplicated Hilbert space.

Actually, both classically and quantum mechanically, the Lyapunov exponent is a distributed quantity: different starting conditions - or different time intervals after the same starting condition - yield different exponents, they are peaked on
a `typical' value and have large (and rare) deviations around, for example when a classical trajectory grazes a regular region.  If the distribution
is not a delta function, it is referred to as `multifractal'\cite{crisanti1988generalized}.

A possibility to study the full distribution of Lyapunov exponents is to introduce the generalized Lyapunov exponents (GLE) $L_{2q}$, defined from the moments of the distribution~\footnote{This may lead to confusion. The Lyapunov exponents are classified as {\em first (maximal), second, ...} 
according to whether they measure linear, area, and in general $k$-form expansions \cite{vulpiani2009chaos}. Each is also generalized according to the moment $q$ considered. Here we are considering expansions of linear lengths, and all moments $q$ of the `first, maximum, Lyapunov exponent', thus $L_{2q}^{(k)}$ for all $q$ and $k=1$}. It turns out that the \emph{quantum} generalized exponents also satisfy themselves a bound, stated in Ref.\cite{tsuji2018bound}, that generalizes Eq.\eqref{eq:bL}. 
These bounds are the subject of this paper.
We have two motivations: Firstly, they put limitations on the chaotic properties of {\em rare} protocols, favouring high and low chaoticity,
and also allow us to define a Lyapunov exponent for {\em typical} protocols, actually different from the usual one considered in quantum mechanics. Secondly, a system that
approaches the bounds for all the $L_{2q}$ will turn out to be {\em mono-fractal}, i.e, the large deviation function becomes peaked on a single value, a property we find intriguing.

The quantum generalized Lyapunov exponents are defined by  considering the $2q$-th commutator between two operators at different times, and assuming that they scale exponentially with time  as
\begin{equation}
	\label{eq:q-comm}
	G_{2q}(t) = 
	\langle \left ( i[A(t), B)] \right ) ^{2q}\rangle_\beta \sim\epsilon ^{2q} \; e^{L^{(\beta)}_{2q} t} \ ,
\end{equation}
where $\epsilon$ is a small parameter, two common examples are   $\epsilon = \hbar$ (the semiclassical limit) or $\epsilon=N^{-1}$ (as in the SYK model), the latter more relevant here as we shall be interested in thermodynamic models.
The thermal average  $\langle \bullet \rangle_\beta$  at inverse temperature $\beta$ may be defined in various ways, as we shall see below. 

The rate $L^{(\beta)}_{2q}$  now defines the \emph{quantum thermal} generalized Lyapunov exponent  of order $q$. The usual (quenched) Lyapunov exponent thus is
\begin{eqnarray}
\label{qle}
\lambda^{(\beta)}_1 = \lim _{q\to 0}  \; \frac{L^{(\beta)}_{2q} }{2q} \ 
\end{eqnarray}
On the other hand, the grow-rate of the square-commutator $\lambda_L$ actually corresponds to the GLE with $q=1$: $\lambda_L = L^{(\beta)}_{2}$.

In the quantum realm, the exponential regime holds only at intermediate times, up to the so-called \emph{Ehrenfest time} $t_{\text{Ehr}}\sim \ln \epsilon^{-1}$. In actual fact, only when $\lim t_{\text{Ehr}} \rightarrow \infty$  is the Lyapunov regime unambiguously defined.

Different $q$-Lyapunovs are dominated by initial conditions having different expansion rates, with larger rates dominating the averages corresponding to larger $q$. Hence, one
must allow for the dependence on the Ehrenfest time itself:
\begin{equation}
    \label{eq:Ehr}
	t^{(2q)}_{\text{Ehr}} = \frac{2q}{L_{2q}}\ln \epsilon^{-1} \quad 
\text{such that} \quad G_{2q}(t)\sim e^{L^{(\beta)}_{2q}(t-t^{(2q)}_{\text{Ehr}})} \ .
\end{equation}
A crucial assumption we shall make here is that:
\begin{equation}
    \label{order}
    t^{(2q)}_{\text{Ehr}} \le t^{(2q')}_{\text{Ehr}} \hspace{1cm} {\mbox{for}} \hspace{1cm} q>q'
\end{equation}
Under these assumptions, we show that the following holds 
\begin{equation}
    \label{eq:BOUNDQ}
	\frac{ L^{(\beta)}_{2q}}{2 q} \leq    \frac{\pi}{\beta \hbar} \ .
\end{equation}
This bound was already stated in Ref.\cite{tsuji2018bound}, without the identification of the rate $ L^{(\beta)}_{2q}$ as a quantum GLE and the relation to large deviations.  We will also clarify some assumptions on which the derivation \cite{tsuji2018bound}.\\
Because we are assuming that $	\frac{ L^{(\beta)}_{2q}}{2 q}$ does not decrease,
the bounds for larger $q$ (but always of order one with respect to $N$) are more stringent. The meaning of this, as the classical discussion below will make clear, is that even rarely expanding conditions are bounded.

\section{Classical generalized Lyapunov exponents}

Let us briefly review the classical case \cite{vulpiani2009chaos}.
 Consider the infinitesimal separation  $|{\bf{\Delta}}(t)|$ between two trajectories at time $t$, starting at a point ${\bf x_0}$ and ${\bf x_0 + \Delta}$:
\begin{equation}
	R( t) = \frac{|{\bf{\Delta}}(t)|}{|{\bf{\Delta}}(0)|} \sim  e^{\lambda \, t} \ .
\end{equation}
The rate $\lambda$ is a function of the initial condition ${\bf x_0}$.
This quantity grows in time according to an asymptotic exponential law $\lambda t$  with $(\lambda>0)$ if the system is chaotic, where $\lambda$ is a function of time which reaches a finite limit at long times.
A very long chaotic trajectory will have explored most of the phase space, and the exponential expansion will be sampled 
from all regions: the value $\lambda$ then becomes essentially the same for all initial conditions. This fact is encompassed in the Oseledec theorem \cite{oseledets1968multiplicative}.

The fact that $\ln R(t)$  is  a cumulative process over stretches of time with uncorrelated properties  leads to
the usual argument for the introduction of a large deviation principle, in this case for
the probability of a Lyapunov value given a random initial condition:
\begin{equation}
\label{cramer}
\ln P(\lambda,t)  \sim - t \; S(\lambda)     \ ,
\end{equation}
where $S$ is the large-deviation (Cram\'er) function. If the system is ergodic, the ensemble of initial conditions may be substituted by the 
ensemble of initial times along the same trajectory.

 The typical Lyapunov exponent $\lambda_1$ is given by
\begin{equation}
	\lambda_1 \equiv \lim_{t\to\infty}\frac 1t \langle \ln R( t) \rangle_\tau \equiv \lim_{t\to\infty}\int d\lambda \; P(\lambda,t) \lambda
\end{equation}
(note that $\lambda=\lambda(t)$).
 
 One can study the $2q$-th moments which, for long enough times, shall grow exponentially as
\begin{equation}
	\label{eq:defLR}
	\mathcal R_{2q}(t) = \langle R( t) ^{2q} \rangle \sim e^{L_{2q} t}  ,
\end{equation}
where
\begin{equation}
	\label{eq:defLC}
	L_{2q} = \lim_{t\to \infty} \frac 1 t \ln \langle R(\tau) ^{2q} \rangle
\end{equation}
are called {the generalized Lyapunov exponents} (GLE) of order $2q$, and characterize the fluctuations of the dynamical system \cite{crisanti1988generalized}, see also Ref.\cite{benzi1984multifractal}. They are defined as \emph{annealed averages}
\begin{equation}
    \label{eq:Lege}
	L_{2q} \equiv \frac 1 t \ln \lim_{t\to\infty}\int d\lambda \; P(\lambda,t) \; e^{2q\lambda t} \sim \max_\lambda  \left\{ - S(\lambda) + 2q \lambda\right\} \ ,
\end{equation}
where we have evaluated the integral over large $t$ using saddle point.
The  typical Lyapunov exponent is retrieved as the limit
\begin{equation}
	\lambda_1 = \lim_{q\to 0} \frac{L_{2q}}{2q} = \frac 12 \frac{dL_{2q}}{dq} \Big |_{q=0}\ .
\end{equation}

The GLE $L_{2q}$ is the Legendre transform of $S(\lambda)$, via Eq.\eqref{eq:Lege}. As such, also $L_{2q}$ is a convex function of $2q$. Using the property of convex differentiable functions \footnote{A convex function $f(x)$ obeys $f(x) \geq f(y) + f'(y) (x-y) $. 
Using this property for $x=2q$ and $y=0$, one has Eq.\eqref{eq_increasingClass}. Using the property for $x=0$ and $y=2q$, one has Eq.\eqref{increasing}.} one obtains that $L_{2q}$ obeys two important properties:
\begin{enumerate}
    \item $L_{2q}/2q$ are an increasing function of the order $q$
    \begin{equation}
    \label{increasing}
	\frac d{dq} \left ( \frac{L_{2q}}{2q} \right ) \geq 0 \ ;
    \end{equation}
    \item the GLEs are always bounded by the linear behaviour
    \begin{equation}
     \label{eq_increasingClass}
	L_{2q} \geq  2q \lambda_1   \ .\\   
    \end{equation}
\end{enumerate}

The equality $L_{2q}=2q \lambda_1$ in Eq.\eqref{eq_increasingClass} holds only if $P(\lambda) = \delta(\lambda-\lambda_1)$, namely if the Lyapunov exponent is the same and does not fluctuate, we have  \emph{mono-fractality}. Another interesting example is the one of random matrices of dimension $D$ without any structure and with high connectivity, which satisfies $L_{2q}= 2q \lambda_1  + \mathcal O(1/D)$ \cite{crisanti1988generalized}. Otherwise, the system is characterized by \emph{multifractal behaviour}.
The higher the moments, the more important the contributions coming from the tales of the distribution. In particular, in the case of a distribution $P(\lambda)$ with a finite support, the limits 
\begin{equation}    
    \label{max}
	\lambda_{max/min} = \lim _{q\to \pm \infty} \frac{L_{2q}}{2q} 
\end{equation}
select the maximal and minimal expanding rates.

\section{Quantum generalized Lyapunov exponents at infinite temperature}
Our goal is to extend the definition of generalized Lyapunov exponents to the quantum domain to discuss the bound in Eq.\eqref{eq:BOUNDQ}.
Systems with a few degrees of freedom do not lend themselves to the implementation of bounds that depend on temperature, as the canonical ensemble is not particularly useful for them. However, we may understand some other features that are also valid in thermodynamic systems by studying infinite-temperature systems of this kind.
In this Section, we define the quantum generalized Lyapunov exponents at infinite temperature and discuss their ``convexity'' properties. We will see that it is straightforward to interpret the quantum GLE as a probe of the spectral properties of the square-commutator operator.

\subsection{Properties of the infinite temperature quantum GLE}
\label{sec:Tinfi}
Let us first analyze the infinite temperature $2q$-th commutator Eq.\eqref{eq:q-comm} at infinite temperature
\begin{equation}
	\label{eq:beta0}
	G^{(0)}_{2q}(t) = \frac {(-1)^q }Z \, \text{Tr} \left (  [A(t), B]  ^{2q} \right ),
\end{equation}
where $Z=\text {Tr}(\mathbb 1)=\dim \mathcal H$ is given by the Hilbert space dimension. This object generalizes the infinite temperature square-commutator in Eq.\eqref{sq}, which has been discussed in a variety of models and it is particularly relevant for dynamical protocols where energy is not conserved (like with periodic driving or in the open system's scenario). 

The infinite temperature quantum GLEs are then defined by the exponential growth at intermediate times
\begin{equation}
	\label{eq:Lbeta0}
	G^{(0)}_{2q}(t) \sim \epsilon^{2q} e^{L^{(0)}_{2q} t} \ .
\end{equation}

We now show that $L^{(0)}_{2q}$ obey the same properties as the classical ones (e.g. Eq.\eqref{increasing}-\eqref{increasing}).
Let us re-write Eq.\eqref{eq:beta0} as
\begin{equation}
	\label{eq:beta0_}
	G^{(0)}_{2q}(t)  = \frac 1Z \sum_i (g_i(t))^{2q} \ ,
\end{equation}
where we have defined $g_i(t)$ the eigenvalues of the square-commutator operator, i.e.
\begin{equation}
    \label{sc_ope}
	- [A(t), B] ^ 2= \sum_{i_t} ({g_i(t)})^2 |i_t\rangle \langle i_t| \ .
\end{equation}
and we have made explicit a factor $t$ so that $\lambda_t$ may have a finite limit. Some properties of this operator have been studied on specific models, see e.g. Refs.\cite{rozenbaum2019universal, gharibyan2019quantum}. If the expectation value of the square-commutator grows exponentially (we consider only times before $t_{\text{Ehr}}$), then it is convenient to write each eigenvalue  as
\begin{equation}
	 g_i(t) =  e^{ \lambda^i_t t} \ .
\end{equation}
By using $1= \int d\lambda \delta(\lambda - \lambda^i_t)$, we can re-write Eq.\eqref{eq:beta0} as
\begin{equation}
\label{23344}
G^{(0)}_{2q}(t) = \int d\lambda\,  P(\lambda) e^{2q \lambda t} 
\end{equation}
where we have defined the \emph{distribution of the quantum local Lyapunov exponents}
\begin{equation}
	\label{eq:Pl_0}
	P(\lambda) = \sum_i\delta(\lambda_{i_t}-\lambda) \ .
\end{equation}
Eq.\eqref{23344}-\eqref{eq:Pl_0} show that the $ G^{(0)}_{2q}(t)$ are moments since they can be written as an integral of times the powers of a function times a positive function $P(\lambda)$. We can associate to the latter a convex Cram\'er function $t S(\lambda) \sim \ln P(\lambda)$ as in Eq.\eqref{cramer}, which gives the Legendre transform of $L^{(0)}_{2q}$.
These relations imply the convexity of the quantum GLE at infinite temperature, which results in:
\begin{enumerate}
\label{properties_beta0}
    \item $ L^{(0)}_{2q}/2q$ is an increasing function of $q$;
    \item the following inequality holds
\begin{equation}
    \label{3Ltilde_0}
	 L^{(0)}_{2q}  \geq 2 q  \lambda^{(0)}_1\ ,
\end{equation}
 where $ \lambda^{(0)}_1 = \lim_{q\to 0} \frac{ L^{(0)}_{2q}}{2q}$. 
\end{enumerate}

The equality holds in the absence of fluctuations in the spectrum of the square-commutator operator. Such mono-fractal behaviour means that -- for the appropriate time's range -- the square-commutator operator is close to a constant times the identity matrix.

\subsection{A semi-classical example: the quantum kicked top}
\label{sec:KT}

As an illustrative example, we study a driven model: the quantum kicked top. Since the energy is not conserved, this model is equivalent to a system at infinite temperature. We thus show that $L^{(0)}_{2q}$ satisfies the properties of convexity and of large deviation theory. (In this section we denote $L^{(0)}_{2q}=L_{2q}$, for the sake of clarity.)

The model is described by the time-dependent Hamiltonian 
\begin{equation}
H = \alpha S_x + \frac{J}N S^2_z \sum_{n=-\infty}^\infty \delta(t - n \tau)\ ,
\end{equation}
where $S_{x, y, z} = \frac 12 \sum_{i=1}^N \sigma^{i}_{x,y,z}$ are collective spin operators.
Due to the collective nature of the interactions, for large $N$ the classical limit is approached. One can define an effective Planck constant
	\begin{equation}
	\label{hbar}
\hbar = \frac 1S = \frac{2}{N}
\end{equation}
that vanishes in the thermodynamic limit. The stroboscopic time-evolution operator (namely, the time-evolution operator over one period) reads
 \begin{equation}
 	\hat U = \hat U _J \hat U_\alpha \, \text{with} \quad  U_\alpha = e^{-i\alpha S_x} \ , \, \hat U_J = e^{-i \frac J N S_z^2} \ .
 \end{equation}
  We fix $\tau=1$  and $\alpha=\pi/2$. Changing the value of the kicking strength $J$, this model undergoes a transition between a regular regime and a chaotic one \cite{haake2010quantum, haake1987classical}. 
  The dynamics of the square-commutator \eqref{sq} have been extensively explored \cite{pappalardi2018scrambling, seshadri2018tripartite, pilatowsky2020positive, sieberer2019digital, lerose2020bridging}. \\

  We consider the strongly chaotic limit by choosing $J=3.5$ and we look at the infinite temperature state with $\rho = \mathbb 1/\dim H$, with $\dim H = N+1$. 
  We study the dynamics of the $G_{2q}(t)$  via exact numerical calculations, specifically via exact diagonalization.  
  We compute the stroboscopic time-evolution of he $2q$-th commutator \eqref{eq:q-comm} using $\hat A = \hat B = \hat S^z$ at times $t = n\tau=0, 1, 2, \dots$.

  \begin{figure}[t]
\begin{center}
\includegraphics[width=.6\columnwidth]{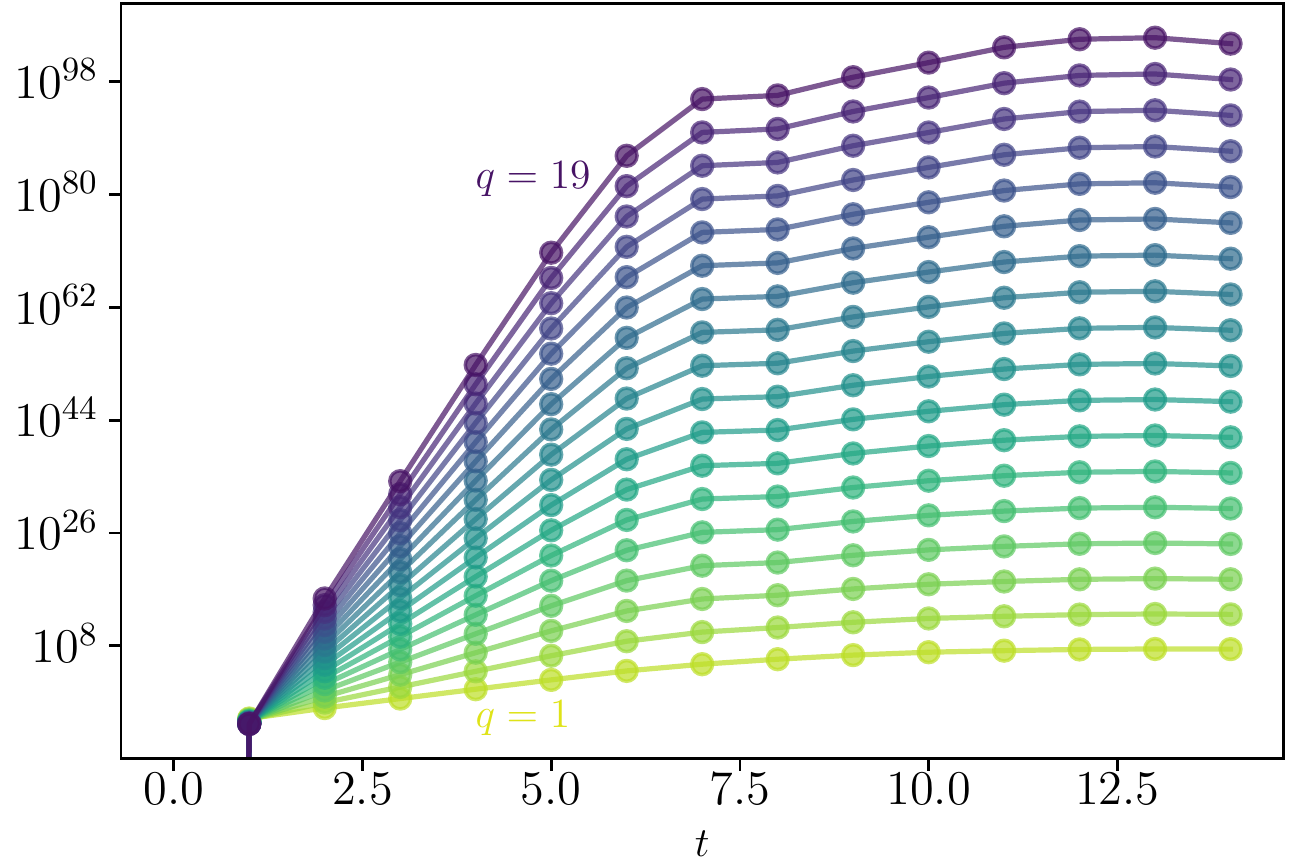}
\caption{Dynamics of $G_{2q}(t)$ in Eq.\eqref{eq:q-comm} for different values of $q=1\div 19$ as a function of time for $N=1600$. }
\label{fig:G2q}
\end{center}
\end{figure}

\begin{figure}[b]
\begin{center}
\includegraphics[width=.6\columnwidth]{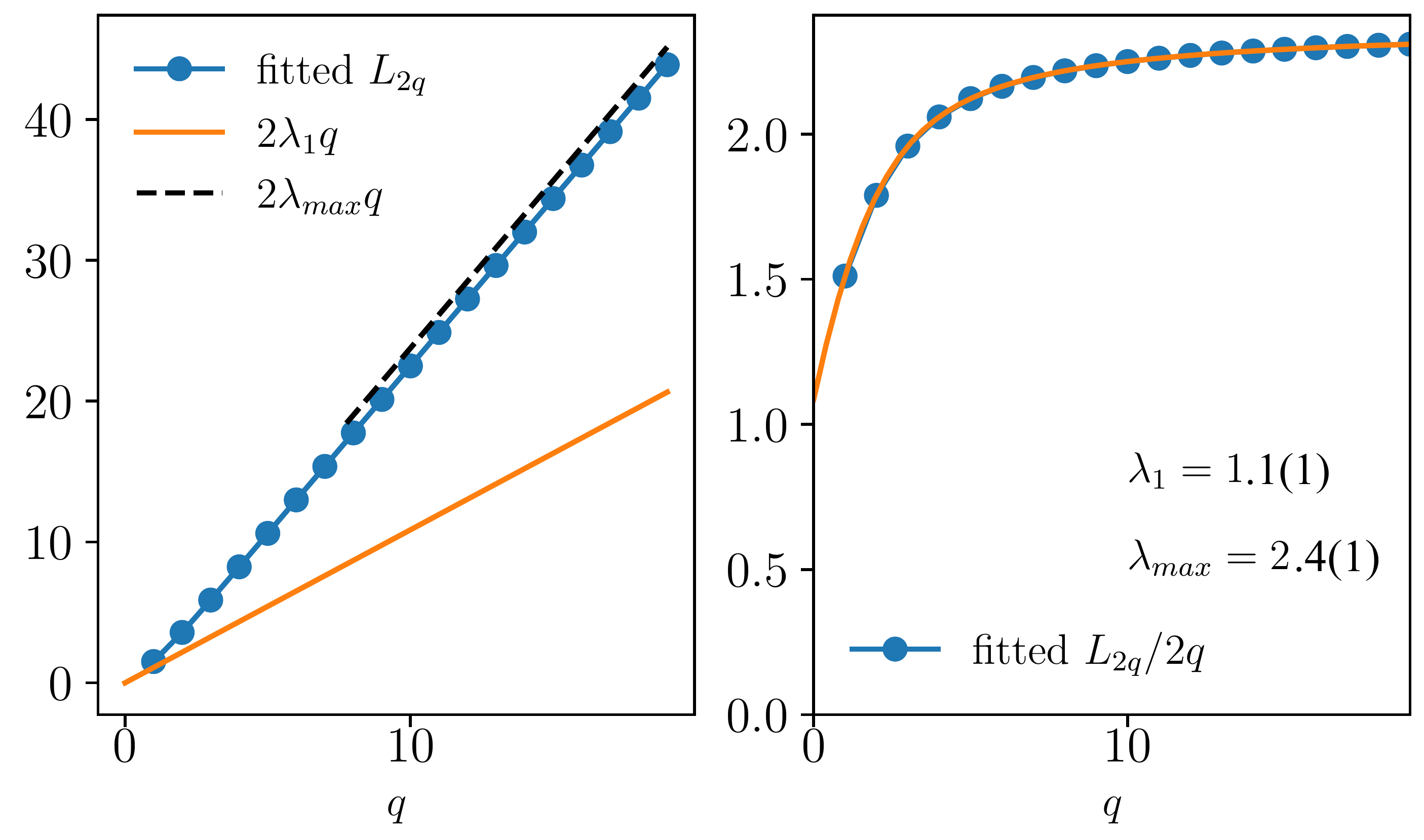}
\caption{Generalized Lyapunov exponents fitted from Fig.\eqref{fig:G2q}. (Left) $L_{2q}$ as a function of $q$, contrasted with the actual Lyapunov exponent $\lambda_1$ and the maximal expanding rate $\lambda_{max}$, obtained by a fit of these data at large $q$. (Right) $L_{2q}/2q$ as a function of the moment $q$, from which we extract the maximal Lyapunov exponent $\lambda_1$.  }
\label{fig:L2q}
\end{center}
\end{figure}

 In Fig.\ref{fig:G2q}, we show the dynamics of the $G_{2q}(t)$ for different values of $q=1\div 19$. The correlators are rescaled by $\hbar^{2q}$ [with $\hbar=2/N$] to emphasise the scaling of Eq.\eqref{eq:q-comm}. Each commutator grows exponentially before the Ehrenfest times with a different rate, that corresponds to the quantum GLE $L^{}_{2q}$. The value of $L_{2q}$ is then fitted and plotted in Fig. \ref{fig:L2q}a, where we display its behaviour as a function of $q$. It is a convex function of $q$ that satisfies $L_{2q} > 2q \lambda_1$ [cf. Eq.\eqref{3Ltilde_0}], being therefore multifractal. The typical Lyapunov exponent $\lambda_1$ is computed as in Fig. \ref{fig:L2q}b, where $L_{2q}/2q$ is plotted as a function of $q$. The extrapolation to $q\to0$ yields $\lambda_1=1.1(1)$, which corresponds to the maximum Lyapunov exponent of the classical model in the chaotic phase  $\lambda_{class}=1.12$, as computed via the Benettin et al. algorithm \cite{Benettin1980P1, Benettin1980P2}, see e.g. the appendix of Ref.\cite{lerose2020bridging}. We also extract the maximal expanding rate $\lambda_{max}=2.4(1)$ [cf. Eq.\eqref{max}] from the limit $q\to\infty$, signalling that the distribution of Lyapunov has finite support.

\begin{figure}[t]
\begin{center}
\includegraphics[width= .6 \columnwidth]{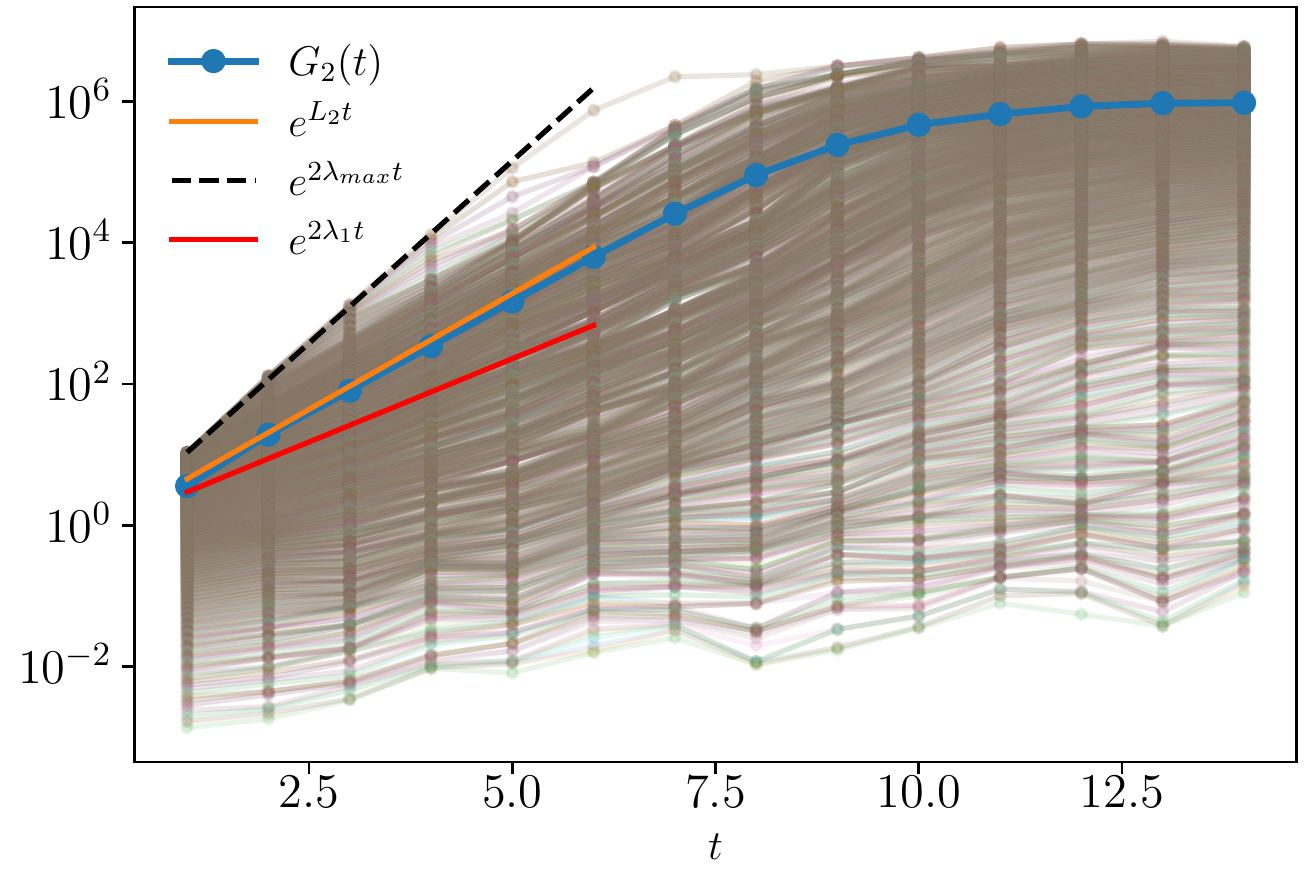}
\caption{Spectrum of the square-commutator $g^2_i(t)$, compared with $G_2(t)$ and the exponential growth with $L_2$, with the maximal expansion rate $\lambda_{max}$ and with $\lambda_1$ for $N=1600$. }
\label{fig:spec}
\end{center}
\end{figure}

\begin{figure}[t]
\begin{center}
\hspace{-2cm}
\includegraphics[width = .54 \columnwidth]{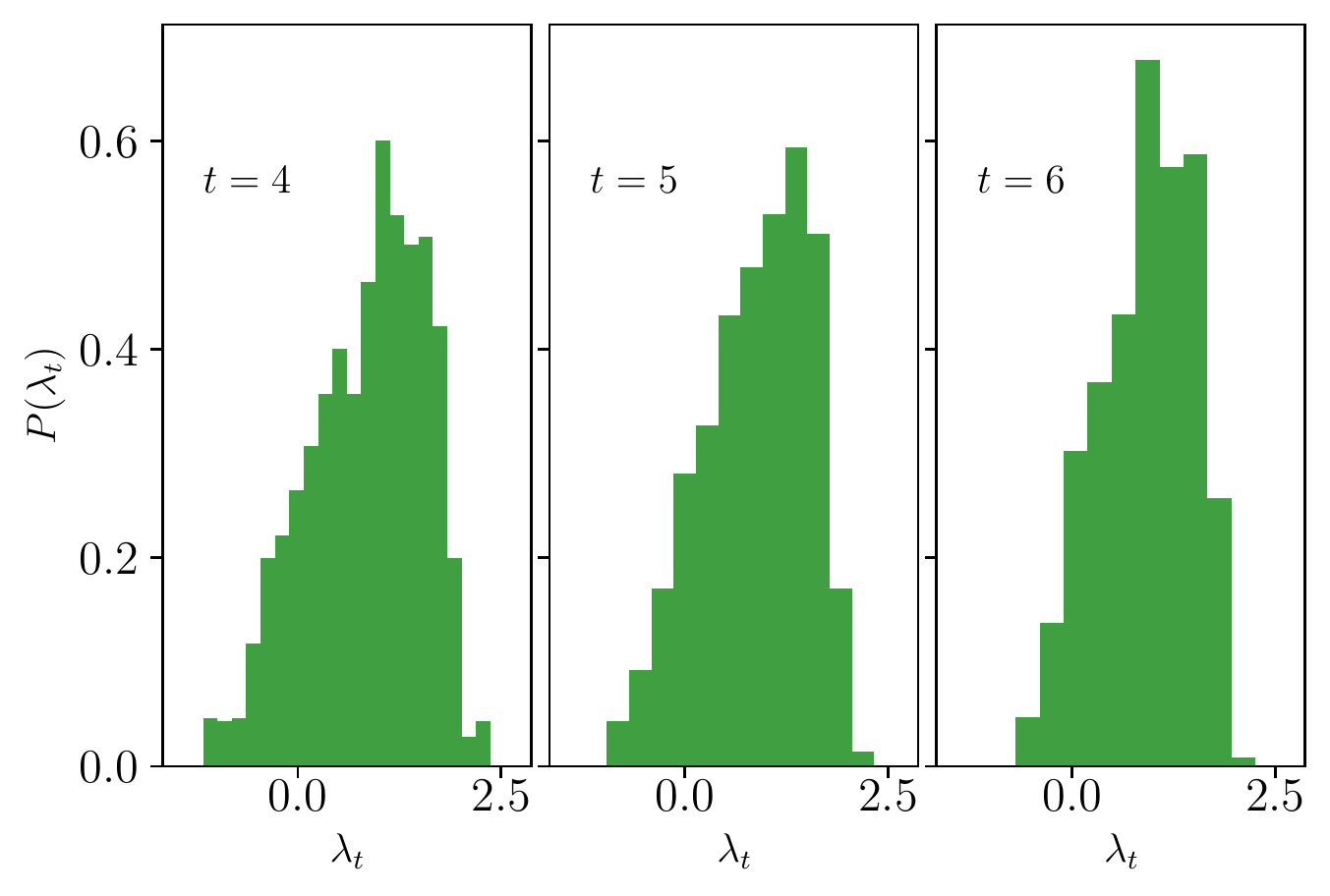}
\includegraphics[width= .4 \columnwidth]{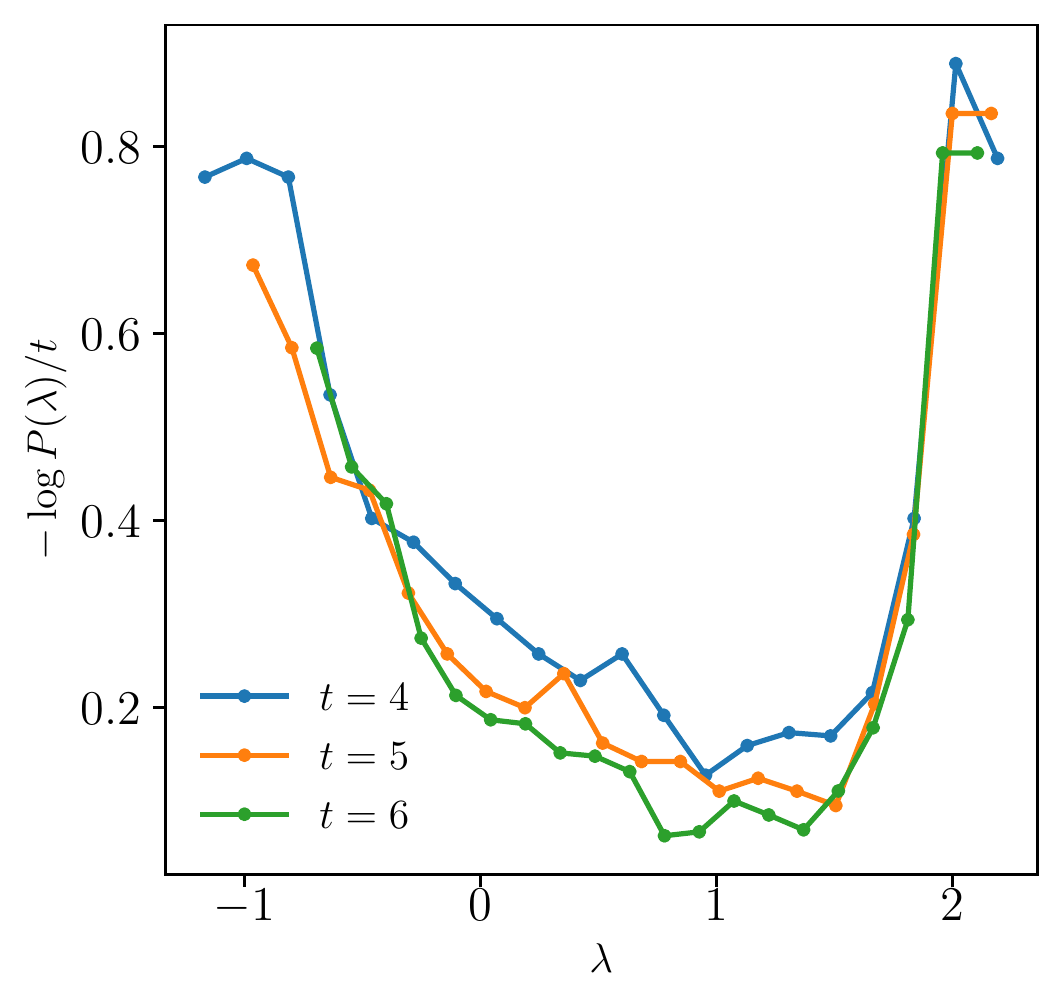}
\caption{Large deviation properties of the spectrum of the square-commutator for $N=1600$. (Left) Numerical distributions of $\lambda^i_t=\ln (g_i(t))/2t$ with $g^2_i(t)$ the eigenvalues of Eq.\eqref{eq:beta0_} at different times $t=3,4,5$. (Right) $- \ln P(\lambda)/t$ with $P(\lambda)$ the empirical distribution. }
\label{fig:LargeDevi}
\end{center}
\end{figure}

In Fig.\ref{fig:spec} (shaded lines) the spectrum of the square commutator in Eq.\eqref{sc_ope}. Most of the eigenvalues grow exponentially in time before saturation and thus define some local Lyapunov exponents.  We compare this behaviour with the standard square-commutator expectation $G_2(t)$ [cf. Eq.\eqref{sq}] (blue dots) which grows exponentially a rate $L_{2}$ larger than the maximum Lyapunov $\lambda_1$. The figure also shows that the $\lambda_{max}$ as fitted extracted in Fig.\ref{fig:L2q} (the dashed black line) corresponds to the maximal expanding rate of the local Lyapunov exponents.

In Fig.\ref{fig:LargeDevi}, we show that the local Lyapunov exponents are a large deviation.  We consider the coefficients $\lambda^i_t=\ln (g_i)/2t$ from Fig.\ref{fig:spec} (we divide everything by a constant factor). On the left, we plot their numerical distribution at different times $t=3,4,5$, which shows that converges to a distribution at large $t$. On the right, we plot $- \ln P(\lambda)/t$ which shall correspond to the smooth convex ``Cramer'' function at large times, see Eq.\eqref{cramer}.

\section{Thermal quantum Generalized exponents}

\label{secAAA}
It is useful to consider the different \emph{regularizations} of the $2q$-th commutator as
\begin{subequations}
	\label{eq:GREGU}
	\begin{align}
	\label{eq:GREGUc}
		G_{2q}^{(\beta)}(t)  & = \text{Tr} \left ( \left (\rho^{\frac 1{4q} } i [A(t), B]\rho^{\frac 1{4q} }
		\right )^{2q}\right )\ ,
		\\
		\label{eq:GREGUf}
		\overline G_{2q}^{(\beta)}(t)  & = \text{Tr} \left ( \left (i [\rho^{\frac 1{8q} } A(t)\rho^{\frac 1{8q} } , \rho^{\frac 1{8q} } B \rho^{\frac 1{8q} } ]
		\right )^{2q}\right ) \ .
	\end{align}
\end{subequations}
with 
\begin{equation}
\rho = e^{- \beta H}/Z_\beta \hspace{1cm} \text{and} \hspace{1cm}Z_\beta = \text{Tr}e^{-\beta H}
\end{equation}
Considering $q=1$ in Eqs.\eqref{eq:GREGU} we retrieve the standard regularized square-commutators for which the bounds have been proved in Ref.\cite{pappalardi2022quantum}.
In Section \ref{sec:twoPointMap} below, we show that the multi-time correlation functions appearing in Eq.\eqref{eq:GREGU} can be mapped as two-times functions in a replicated Hilbert space of $2q$ copies. 
This allows one to rationalize the use of such regularizations - that might seem an artificial construction - as fixing the temperature of the different replicas to be the same. \\

These regularizations define the thermal average introduced in Eq.\eqref{eq:q-comm}, that defines the thermal GLE $L^{(2q)}_\beta$.
We consider the situation in which both of the $q$-th commutators grow exponentially in time as 
\begin{align}
	\label{eq:lambdaQQ}
	G^{(\beta)}_{2q}(t) \propto \epsilon^{2q} e^{L_{2q}^{(\beta)} t} \ ,
	\\
	\bar G^{(\beta)}_{2q}(t) \propto\epsilon^{2q} e^{L_{2q}^{(\beta)} t} \ ,
\end{align}
which is valid only for an intermediate time regime 
\begin{equation}
    t_d \ll t \leq t^{(2q)}_{\text{Ehr}}\equiv \frac{L^{(\beta)}_{2q}}{2q} \ln \epsilon^{-1} \ .
\end{equation}

\subsection{From commutators to OTOCS}
\label{secOtoc}

Consider the quantities of Equation (\ref{eq:GREGU}). Expanding the commutators, we get a series of OTOC terms containing exactly $k$ times
\begin{equation}
   G_{2q}^{(\beta)}(t) = \sum_{k=1}^{2q} d_k \, [OTOC]_{k} 
   \end{equation}
   of the form
   \begin{equation}
     [OTOC]_{k} = \langle A_1(t) B_1(0)  \dots A_{k}(t) B_{k}(0) \rangle_\beta  + \text{h.c.} \ ,
\end{equation}
where $A_1, B_1, ...$ are powers of $A$ and $B$ and and $d_k$ some coefficients. With these notations, $[OTOC]_{1}=\langle A^q(t)B^q(0)\rangle_\beta + \text{h.c.}$ is a function of two times.  
Out-of-time-order correlators between $k$ operators are sometimes referred to as $k$-OTOC, see e.g. Refs.\cite{roberts2017chaos,  haehl2018effective, haehl2018fine, bhattacharyya2022towards}.
We are here interested in understanding their structure in time for exponential growth.
Following \cite{maldacena2016bound}, we assume that there exists some dissipation time $t_d$, after which two-point functions factorize as $[OTOC]_{1}\sim C_q$. 

Each $[OTOC]_{2k}$ may encode at most as fast as the corresponding Lyapunov behaviour, during the corresponding Ehrenfest time
\begin{equation}
C_k - [OTOC]_{2k} \propto  \epsilon^{2k} e^{L_{2k} t} = e^{L_{2k} (t-t_{\text{Ehr}}^{(2k)})} \ .
\end{equation}
If we evaluate this term at times corresponding to a finite but small fraction
of the corresponding Ehrenfest time $t_{\text{Ehr}}^{(2q)}$
we conclude that all the terms with $k<q$ are of lower or equal order, because of the ordering of Ehrenfest times [cf. Eq.\eqref{order}].
We thus conclude that
\begin{subequations}
	\label{eq:GREGU1}
	\begin{align}
		G_{2q}^{(\beta)}(t)  & =  \text{Tr} \left ( \left (\rho^{\frac 1{4q} } i [A(t), B]\rho^{\frac 1{4q} }
		\right )^{2q}\right )\ \sim C_q - \Re \; \text {Tr} \left (\, 
	\rho^{\frac 1{2q}} A(t) B\, \rho^{\frac 1{2q}} A(t) B\,  \dots \rho^{\frac 1{2q}} A(t) B\, 
	\right )\ , \nonumber
		\\
		\overline G_{2q}^{(\beta)}(t)  & =  \text{Tr} \left ( \left (i [\rho^{\frac 1{8q} } A(t)\rho^{\frac 1{8q} } , \rho^{\frac 1{8q} } B \rho^{\frac 1{8q} } ]
		\right )^{2q}\right ) \ \sim \overline C_q -\text {Tr} \left (\, 
	\rho^{\frac 1{4q}} A(t)\rho^{\frac 1{4q}} B\, \rho^{\frac 1{4q}} A(t)\rho^{\frac 1{4q}} B\,  \dots \rho^{\frac 1{4q}} A(t)\rho^{\frac 1{4q}} B\, 
	\right )\ , \nonumber
	\end{align}
\end{subequations}
where the constants $C_q, \bar C_q$ are different given the different regularizations.

\subsection{Product space,  Fluctuation-Dissipation Theorem and bound}
In this section, we will show how the multi-time OTOC appearing in the generalized $2q$-th commutators have a simple interpretation as two-times correlation functions in a replicated space. This allows us to easily demonstrate that such OTOC obeys the fluctuation-dissipation theorem (FDT).

\label{sec:twoPointMap}
Let us consider the following $4q$ point out of time order correlator
\begin{align}
	\label{sedS2}
	S_{2q}(t) & = \frac 1Z_\beta \text {Tr} \left (\, 
	(\rho^{\frac 1{2q}} A(t) B\, )^{2q}\, 
	\right ) 	= \text {Tr} \left (\, 
	\rho^{\frac 1{2q}} A(t) B\, \rho^{\frac 1{2q}} A(t) B\,  \dots \rho^{\frac 1{2q}} A(t) B\, 
	\right )\ .
\end{align}
Now, we re-write it in terms of the spectral representation of the Hamiltonian $H|n\rangle = E_n|n\rangle$ as 
\begin{align}
	S_{2q}(t) & =
		\label{eq:7}\frac 1Z_\beta
	 \, \sum_{n_1 n_2 \dots n_{2q}} e^{-\frac \beta {2q}(E_{n_1}+E_{n_2}+\dots E_{n_{2q}})}  
	\quad \times\langle n_1|A(t) B|n_2\rangle  \langle n_2|A(t) B|n_3\rangle\dots  
	\langle n_{2q}|A(t) B|n_1\rangle	\nonumber
	\\
	& = \frac 1Z_\beta \sum_{n_1 n_2 \dots n_{2q}}e^{-\frac \beta {2q}(E_{n_1}+E_{n_2}+\dots E_{n_{2q}})}  \quad \times
	\langle n_1 n_2 \dots n_{2q-1}n_{2q} |\mathbb A(t) \mathbb B |n_2 n_3 \dots n_{2q} n_1\rangle \ , \nonumber
\end{align}
where we introduced the operators that act in the $2q$-th replicated Hilbert space 
\begin{align}
    \mathbb A(t) & = A(t) \otimes A(t) \dots \otimes A(t) \ , \hspace{2cm} 
    \mathbb B  = B \otimes B \dots \otimes B \  ,
\end{align}
and the replicated Hamiltonian \begin{equation} \mathbb H  = H\otimes 1\dots \otimes 1 + 1\otimes H\dots \otimes 1 + \dots + 1\otimes 1\dots \otimes H .
\end{equation} 
We also define the \emph{cyclic shift operator} $\mathbb P$ that permutes cyclically states between the Hilbert spaces as
\begin{align}
	\mathbb P|  n_1 n_2 \dots n_{2q-1}n_{2q} \rangle &= |n_2 n_3 \dots n_{q} n_1\rangle \quad \text{with } \quad \mathbb P^{2q} = 1 \quad \text{,}   \\
	\mathbb P^\dagger |  n_1 n_2 \dots n_{2q-1}n_{2q} \rangle & = |  n_{2q}n_1 n_2 \dots n_{2q-1} \rangle \ .
\end{align}
Notice that the operator $\mathbb P$ is non-hermitian, but we can define
$	\tilde {\mathbb P}  = \frac{\mathbb P + \mathbb P^\dagger}2$ that is.   $	\tilde {\mathbb P}$ also commutes with  $\mathbb A(t)$, $\mathbb B$ and $\mathbb H$, so that
$\tilde {\mathbb P}\mathbb B$
is Hermitian.

Let us re-write Eq.\eqref{eq:7} as
\begin{align}
		\label{eq:13}
	S_{2q}(t) & =
	 \frac 1Z_\beta \sum_{n_1 n_2 \dots n_{2q}} e^{-\frac \beta {2q}(E_{n_1}+E_{n_2}+\dots E_{n_{2q}})}
	 \times
	 \frac 12 \Big (
\langle n_1 n_2 \dots n_{2q-1}n_{2q} |\mathbb A(t) \mathbb B |n_2 n_3 \dots n_{q} n_1\rangle  \nonumber
\\& \quad + 
\langle n_1 n_2 \dots n_{2q-1}n_{2q} |\mathbb A(t) \mathbb B |n_{2q} n_{1} \dots  n_{q-2}n_{2q-1}\rangle 
	\Big )\ ,\nonumber
\end{align}
where in the second line we have simply used a different resolution of the identity and a reshuffling of the matrix elements \footnote{We use
\begin{align*}
	& \langle n_1 |
	A(t) B
	|n_{2q}
	\rangle  
	\langle n_{2q}|A(t) B|n_{2q-1}\rangle\dots  
	\langle n_3|A(t) B|n_2\rangle
 \langle n_2|A(t) B|n_1\rangle
\\  
& \quad =
 \langle n_1|A(t) B|n_{2q}\rangle \langle n_2|A(t) B|n_1\rangle  
\\ 
&  \quad \quad \times f
  \langle n_3|A(t) B|n_2\rangle \dots \langle n_{2q}|A(t) B|n_{2q-1}\rangle \ .
\end{align*}
}.
Therefore we  can re-write Eq.\eqref{sedS2} as
\begin{equation}
    \label{replicawonders}
	S_{2q} (t) = \frac 1Z_\beta \text{Tr} \left ( e^{-\beta_{2q} \mathbb H} \, \mathbb A(t) \mathbb B \, \tilde{\mathbb P}
	\right )
\end{equation}
which, besides a normalization, is a standard equilibrium expectation value of a two-time function at inverse temperature $\beta_{2q} = \beta /2q$. This result naturally generalizes the one for four times OTOC, for which $\mathbb P = \mathbb P^\dagger = \tilde{\mathbb P}$, 
as derived in Ref.\cite{pappalardi2022quantum}.

\paragraph{Fluctuation-dissipation in  the replicated space}
We may now write the extended KMS relations. We consider
\begin{subequations}
\label{eq:CR}
\begin{align}
	{\mathbb C}_{2q}(t)  =&
    \frac 12 \frac 1Z\text{Tr} \left [ {e^{-\beta_{2q} \mathbb H} }\,\{ \mathbb A(t) , \, \mathbb B\, \tilde {\mathbb P}\}	\right ]  \ ,
    \\
	{\mathbb R}_{2q}(t)  =& \frac i \hbar \theta(t) \frac 1Z \text{Tr} \left [ {e^{-\beta_{2q} \mathbb H} }\, [\mathbb A(t) ,\, \mathbb B\, \tilde{\mathbb P}]	\right ] \\
	{\mathbb F}_{2q}(t)  =&  \frac 1Z\text{Tr} \left [ {e^{-\frac{\beta_{2q}}{2} \mathbb H} }\,\mathbb A(t)    {e^{-\frac{\beta_{2q}}{2} \mathbb H} }     \, \mathbb B\, \tilde {\mathbb P}	\right]
	\label{regulaF}
\end{align}    
\end{subequations}
where  $\mathbb C_{2q}$ and $\mathbb R_{2q}$ are defined as usual from real and  imaginary parts of $S_{2q}(t) = \mathbb C_{2q}(t) + \hbar (\mathbb R_{2q})''(t)$ and correspond to fluctuations and response functions respectively. Instead, the (Whiteman) correlation function $\mathbb F_{2q}$ in the original space is
\begin{equation}
	\label{Fgrasso}
    \mathbb F_{2q}(t) = \text {Tr} \left ( ( \rho^{1/4q} A(t)  \rho^{1/4q} B ) ^{2q}
    \right ) \ .
\end{equation}
We remark that the Fourier transforms of the connected parts of $\mathbb F_{2q}$, known as free cumulants, directly encode the energy shell correlations appearing in the eigenstate thermalization hypothesis \cite{foini2019eigenstate, pappalardi2022eigenstate}.

The correlation functions defined in Eq.\eqref{eq:CR} obey the FDT at a modified temperature $\beta_{2q}$ \cite{tsuji2018out}. In the frequency domain, the FDT reads
\begin{subequations}
\begin{align}
\label{eq:fdtC}
	\mathbb C_{2q}(\omega) & = \cosh(\beta_{2q} \hbar \omega /2)  \mathbb F_{2q}(\omega) 
		\ ,\\
\label{eq:fdtR}
	\hbar ({\mathbb R_{2q}})''(\omega) & = \sinh(\beta_{2q} \hbar \omega /2)  \mathbb F_{2q}(\omega)
		\ ,
\end{align}
\end{subequations}
equivalent to the standard formulation $\hbar  (\mathbb R_{2q})''(\omega) =  \tanh(\beta_{2q} \hbar \omega /2) C_{2q}(\omega)$. We are interested in correlations in the time domain, hence at the fluctuation-dissipation theorem formulated in the time domain, the $t$-FDT \cite{pappalardi2022quantum}. In particular, we will use the following relations
\begin{subequations}
\label{tfdt}
\begin{align}
\label{eq:TfdtC}
	\mathbb C_{2q}(t) & = \cos \left (\frac{\beta_{2q} \hbar}2 \frac{d}{dt} \right )  \mathbb F_{2q}(t) 
		\ ,\\
\label{eq:tfdtR}
	\hbar ({\mathbb R_{2q}})''(t) & = \sin \left ( \frac {\beta_{2q} \hbar} 2 \frac{d}{dt} \right )  \mathbb F_{2q}(t)
		\ .
\end{align}
\end{subequations}

\paragraph{The bound}
At times small but comparable with $t_{\text{Ehr}}^{(2q)}$, the previous arguments showed that the $2q$-OTOC are dominated by the regularized commutators  [cf. Eq.\eqref{eq:GREGU1}]
\[
  C_q - \mathbb C_{2q}(t)\sim  G_{2q}^{(\beta)}(t) \ , \quad \quad 
 \bar C_q - \mathbb F_{2q}(t) \sim   \bar G_{2q}^{(\beta)}(t) 
\]
when the behavior is exponential as $\sim \exp[{L^{(\beta)}_{2q}(t-t^{(2q)}_{\text{Ehr}})}]$. The $t$-FDT in Eq.\eqref{eq:TfdtC} implies
\begin{equation}
\label{cosine}
\frac{C_q -\mathbb C_{2q}(t)}{\bar C_q - \mathbb F_{2q}(t)}= \cos \left ( \frac{\beta_{2q} \hbar L^{(\beta)}_{2q}}{2} \right)
\end{equation}


The positivity of these coefficients -- that follows from the fact that the $2q$-th commutators are positive definite --  requires that the GLE must be such that $ {\cos \left ( \frac{\beta_{2q} \hbar  }{2} L^{(\beta)}_{2q}\right )\geq 0}$. We thus conclude 
\begin{equation}
    \label{babound}
	 \frac{L^{(\beta)}_{2q} }{2 q}\leq    \frac{\pi}{\beta \hbar} \ .
\end{equation}
In the models where the Lyapunov depends on temperature \cite{maldacena2016remarks, Kurchan2018Quantum, pappalardi2021low}, the cosine above in Eq.\eqref{cosine} starts from zero at large temperature and is always in the first quadrant.

The bound on the $2q$-th OTOC rate was previously derived by Tsuji et al. in Ref.\cite{tsuji2018bound}, by taking Eq.\eqref{eq:GREGU1} as a working assumption. In the section \ref{secOtoc} above, we have justified it using the ordering of the Ehrenfest times $t^{(2q)}_{\text{Ehr}}$.

\subsection{Distribution functions}

The generalized Lyapunov exponents are the moments of a Lyapunov distribution function,
as we have seen in the classical case and for the quantum GLE at infinite temperature. 
In the case of finite $\beta$, the structure is more complex. This is due to the presence of $q-$dependent thermal matrices $\rho$ in the definition of the regularized powers of commutators in Eqs.\eqref{eq:GREGU}.

Nevertheless, one may define the Legendre transform of the thermal GLE as
\begin{equation}
    S(\lambda, \beta) = \max_q(2\lambda q - L^{(\beta)}_{2q} ) \ .
\end{equation}
In analogy with the previous cases, we may interpret it as the Cramèr function of an associated large deviation function $P(\lambda, \beta) \sim \exp(S(\lambda, \beta) t)$. As such, it shall obey similar properties as discussed above. In particular, 
the convexity of $S(\lambda, \beta)$ and $L^{(\beta)}_{2q}$ 
corresponds with the ordering of the Ehrenfest times in Eq.\eqref{order} assumed at the beginning.
The latter is equivalent to the conditions that
\begin{equation}
    \frac{L^{(\beta)}_{2q}}{2q} \quad \text{increasing function of }q.
\end{equation}
It is thus clear, that the quantum bound \eqref{babound} constraints the larger $q$ that are related with the rare large deviations.

\section{Discussion and conclusions}

In this work, we introduced the quantum generalized Lyapunov exponents that quantify the large deviations of the spectrum of an appropriate operator. First, we discussed their convexity properties at infinite temperatures, which we exemplified on the kicked top. At finite temperatures, the quantum fluctuation-dissipation-theorem (KMS) imposes a bound on their value, thus generalizing the celebrated bound to chaos to multipoint correlations. These bounds set a limit on the large deviations of chaotic properties.

A fascinating point is the interpretation of saturating the bound \eqref{eq:BOUNDQ} at every $q$, which implies a form of mono-fractality 
\begin{equation}
    \label{satur}
     L^{(\beta)}_{2q}  = \frac{\pi \hbar}{\beta}\, 2q \ .
\end{equation}
Classical examples of mono-fractal behaviour, {\it i.e. models for which every trajectory has the same Lyapunov exponent}, are the backer map \cite{aizawa1982global} and the free dynamics on the pseudosphere (the surface with constant negative curvature) \cite{balazs1986chaos}.
What can we learn about the models that saturate the quantum bound \eqref{satur}?
A natural expectation is that the SYK model would lie in this class. In this case, it would be interesting to explore the meaning of such quantum mono-fractality in connection to the distinct properties of the model.

\section*{Acknowledgments}
S. P. has received funding from the European Union’s Horizon Europe program under the Marie Sklodowska Curie Action VERMOUTH (Grant No. 101059865).

\bibliographystyle{JHEP}
\bibliography{biblio.bib}

\end{document}